\documentclass[%
 reprint,
 superscriptaddress,
 showpacs,preprintnumbers,
 amsmath,amssymb,
 aps,
pra,
 floatfix,
]{revtex4-1}

\usepackage[T1]{fontenc} 

\usepackage{graphicx}
\usepackage{dcolumn}
\usepackage{bm}
\usepackage[
breaklinks,
]{hyperref}

\begin{document}
\bibliographystyle{apsrev4-1}

\title{Characterization of the stimulated excitation in a driven Bose-Einstein condensate}%

\author{Tao Chen}%
\email{phytch@zju.edu.cn}
\author{Bo Yan}%
\email{yanbohang@zju.edu.cn}
 
\affiliation{%
Department of Physics, State Key Laboratory of Modern Optical Instrumentation, Zhejiang University, Hangzhou, China, 300012
}%
\affiliation{%
 Collaborative Innovation Centre of Advanced Microstructures, Nanjing University, Nanjing, China, 210093
}%



\begin{abstract}

We apply the time-dependent generalized Hartree-Fock-Bogoliubov (td-GHFB) theory to describe the stimulated excitation driven by periodically modulating the interactions in a Bose-Einstein condensate (BEC). A comparison with the results calculated from the typical Bogoliubov approximation indicates that the additional interaction terms contributed by the excited modes play a significant role to explain the dynamics of the stimulating process. The td-GHFB model has not only painted a clear picture of the density wave propagation, but also partly explained the generation of the second order harmonic of the excited modes. The theorectical framework can be directly employed to study similar driven processes.

\end{abstract}
\maketitle


\section{\label{section1}Introduction}

Quantum coherent manipulations of the microscopic atomic processes in many-body systems have revolutionized the development of atomic, molecular and optical physics \cite{Raimond2001,Chin2010,Georgescu2014}. In recent years, pairwise scattering phenomena induced by driving a BEC have attracted lots of attentions \cite{Zin2005,Perrin2007,Sykes2008,Kheruntsyan2012}, from earlier twin matter-wave beam separation \cite{Bucker2011} to recent Bose fireworks \cite{Clark2017}. Observations of these stimulated emissions demonstrate the analogy between matter waves and classical light for further investigations on the fundamental concepts in quantum mechanics, such as quantum correlations and entanglements \cite{Cronin2009,Kwiat1995,Sorensen2001,Amico2008,Horodecki2009,Reid2009,Ross2013}.

In the twin-beam experiment \cite{Bucker2011}, the viewpoint of parametric down-conversion in nonlinear optics can be straightforwardly applied \cite{Wasak2014} since the dynamics of the atom-pairs are governed by the nonlinear terms in the Gross-Pitaevskii (G-P) and Bogoliubov equations. Similarly, the jet emission caused by the periodic modulation of the scattering length should resort to the wave mixing between external magnetic photons and matter waves\cite{Feng2018}, and a time dependent mean-field theory within the Bogoliubov approximation \cite{Wu2018,Fu2018} has been developed to explain the non-equilibrium dynamics in such systems, i.e. the asymmetric angular density correlations. However, two problems of the used Bogoliubov mean-field theory should be addressed: (i) The typical Bogoliubov approximation indicates that the condensate is macroscopically occupied and the excitation of the condensate is negligible, which is valid at the initial stage of the modulation, but collapses after the excited modes become largely populated, especially for a rather long modulating time; (ii) The high-order terms above quadratic are neglected, which makes the high-order collisions between the excited modes disappear, and consequently can not explain the experimentally observed high-order matter-wave harmonics \cite{Feng2018}. To overcome these, one should work with the full expansion of the many-body Hamiltonian, or at least some necessary high-order corrections to the typical Bogoliubov mean-field approximation.

\begin{figure}[b]
\includegraphics[width= 0.5\textwidth]{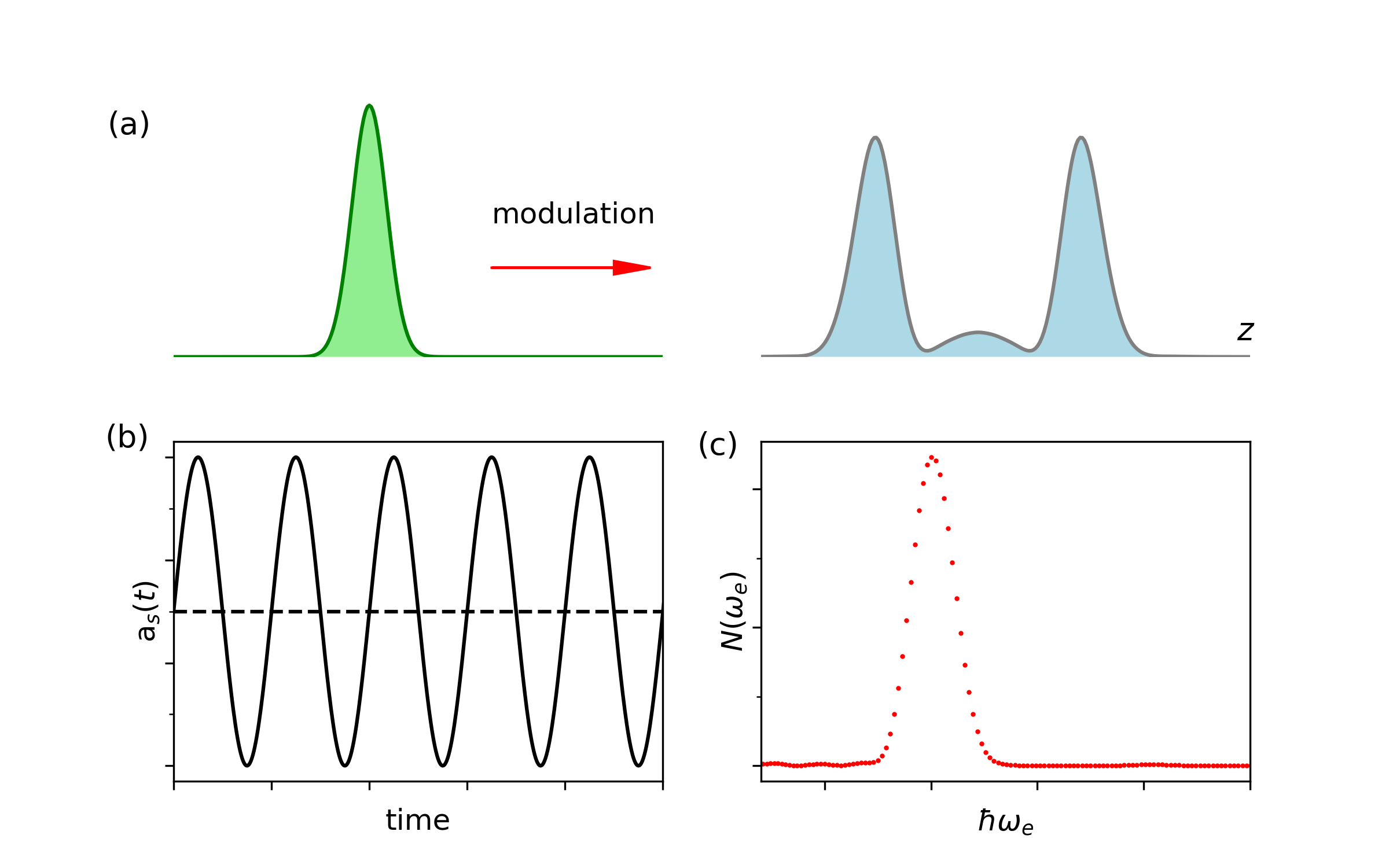}
\caption{\label{fig1} Diagram of the stimulated excitation in a driven BEC. (a) shows the one dimensional density distributions. The BEC gets spatially separated after a modulation, indicating that the stimulated emission happens. (b) shows the time-dependent s-wave scattering length with a modulation amplitude of $a_\text{s,t}$ and a modulation frequency of $\omega$. The dashed line indicates the initial background s-wave scattering length value $a_\text{s,bg}$. (c) The distribution of the population in each excited mode. $N(\omega_\text{e})$ means the excited atom number in the mode with energy $\hbar\omega_\text{e}$. The peak locates around the mode with energy $\hbar\omega/2$ with a quasi-Gaussian shape. A Gaussian fit tells us the full width of half maximum (FWHM).}
\end{figure}

In this work we employ the time-dependent generalized Hartree-Fock-Bogoliubov theory \cite{Griffin1996} to characterize the stimulated excitation. As shown in Fig.\ref{fig1}, by periodically modulating the scattering length near a zero point of a Feshbach resonance, i.e., $a_\text{s}(t) = a_\text{s,bg} + a_\text{st}\text{sin}(\omega t)$, a condensate gets spatially separated and the excited parts accumulate around the mode of $\hbar\omega_\text{e}=\hbar\omega/2$ due to the energy conservation. Different from the typical Bogoliubov approximation, here the contribution of the interactions from the excited fractions has been taken into account to correct the time evolution of the G-P equation and the corresponding Bogoliubov equations. We show that the td-GHFB theory well captures the dynamics of the emission process and clearly explains the interference and the density wave \cite{Staliunas2002}. Furthermore, the td-GHFB theory takes some parts of the high-order interaction terms into consideration, partly explaining the generation of the high-order harmonics with a large modulating amplitude $a_\text{st}$. 

\section{the td-GHFB theory}

We consider a trapped BEC at zero temperature described by the time-dependent many-body Hamiltonian, 
\begin{eqnarray}
H(t) &=& \int d^3 r \hat{\psi}^\dag(r,t) H_0\hat{\psi}(r,t) + \nonumber\\[0.5em]
 ~&~&\frac{g(t)}{2}\int d^3r \hat{\psi}^\dag(r,t) \hat{\psi}^\dag(r,t) \hat{\psi}(r,t)\hat{\psi}(r,t),\label{eq1}
\end{eqnarray}
where $H_0=-\frac{\hbar^2\nabla^2}{2m}+V(r)$, $g(t)=4\pi\hbar^2a_\text{s}(t)/m$, and $V(r)$ is the external trapping potential. The driven dynamics can be obtained from the Heisenberg equation of motion for the field operator $\hat{\psi}$, that is, $i\hbar\partial_t\hat{\psi}=H_0\hat{\psi} + g(t)\hat{\psi}^\dag\hat{\psi}\hat{\psi}$. By decomposing the field operator into a condensate part $\phi$ and an excited fluctuation field operator $\hat{\psi}_\text{e}$ as $\hat{\psi}=\phi + \hat{\psi}_\text{e}$, we get the interaction term $\hat{\psi}^\dag\hat{\psi}\hat{\psi}=|\phi|^2\phi + 2|\phi|^2\hat{\psi}_\text{e} + \phi^2\hat{\psi}^\dag + \phi^*\hat{\psi}_\text{e}\hat{\psi}_\text{e} + 2\phi\hat{\psi}^\dag_\text{e}\hat{\psi}_\text{e} + \hat{\psi}^\dag_\text{e}\hat{\psi}_\text{e}\hat{\psi}_\text{e}$, and the last part can be treated under mean-field approximation, i.e., $\hat{\psi}^\dag_\text{e}\hat{\psi}_\text{e}\hat{\psi}_\text{e}\approx 2 \langle\hat{\psi}^\dag_\text{e}\hat{\psi}_\text{e}\rangle\hat{\psi}_\text{e} + \langle\hat{\psi}_\text{e}\hat{\psi}_\text{e}\rangle\hat{\psi}^\dag_\text{e}$. Then, we obtain the generalized time-dependent G-P equation \cite{Griffin1996}
\begin{equation}
i\hbar\partial_t\phi = H_0\phi + g\left[(n_0 + 2n_\text{e})\phi+m_\text{e}\phi^*\right], \label{eq2}
\end{equation}
where the density of the condensate $n_0(r,t) = |\phi(r,t)|^2$, the density of the excited atoms $n_\text{e}(r,t) = \langle\hat{\psi}^\dag_\text{e}(r,t)\hat{\psi}_\text{e}(r,t)\rangle$ and the excited pairing field $m_\text{e}(r,t) = \langle \hat{\psi}_\text{e}(r,t)\hat{\psi}_\text{e}(r,t)\rangle$. Equation (\ref{eq2}) determines the evolution of the condensate wave-function $\phi(r,t)$, and the contributions from the excited modes are included. The dynamics of the excited modes are yielded by performing the Bogoliubov transformation $\hat{\psi}_\text{e}(r,t) = \sum_j \left(u_j(r,t)\hat{\alpha}_j - v_j^*(r,t)\hat{\alpha}_j^\dag\right)$ [$\hat{\alpha}_j$ and $\hat{\alpha}_j^\dag$ are the quasi-particle operators], resulting that the Bogoliubov amplitudes are determined by \cite{Griffin1996}
\begin{eqnarray}
i\hbar\partial_t u_j &=& (H_0-\mu + 2g(n_0+n_\text{e}))u_j  - g(\phi^2+m_\text{e})v_j, \label{eq3}\\[0.5em]
-i\hbar\partial_t v_j &=& (H_0-\mu + 2g(n_0+n_\text{e}))v_j  - g(\phi^{*2}+m^*_\text{e})u_j, \label{eq4}
\end{eqnarray}
with $\mu$ the initial chemical potential of the condensate, and $n_\text{e}(r,t)=\sum_j |v_j(r,t)|^2$, $m_\text{e}(r,t)=-\sum_j u_j(r,t)v^*_j(r,t)$ \cite{Dalfovo1999}. The coupled equations (\ref{eq2})-(\ref{eq4}) should be numerically solved under the conservation of the particle number $N=\int d^3r \left[n_0(r,t) + n_\text{e}(r,t)\right]$ to obtain the time-dependence of the population in each excited mode. The atom number in the $j$-th excited mode is given by $N(\omega_j)=\int d^3r |v_j(r,t)|^2$. If we set $n_\text{e}$ and $m_\text{e}$ equal to zeros, Eqs.(\ref{eq2})-(\ref{eq4}) then reduce to the usual G-P and Bogoliubov-de Gennes equations in the typical time dependent Bogoliubov theory \cite{Wu2018}. The initial condensate wave-function $\phi(r, t=0)$ can be obtained by numerically solving the stationary G-P equation with the imaginary time backward Euler pseudo-spectral method \cite{Amara1993}, while the $u_j(r,t=0)$ and $v_j(r,t=0)$ series are calculated from the Bogoliubov-de Gennes equations \cite{Dalfovo1999} with the initial scattering length $a_\text{s,bg}$.  The initial excited amplitudes chosen here are different from those in Ref.\cite{Fu2018} where a random fluctuation was employed.

For simplicity, we apply the above td-GHFB theory to investigate a simple quasi one dimensional system with a non-negative background scattering length $a_\text{s,bg}$, and it can be directly extended to study two and three dimensional cases. The trap potential $V(r) = m(\omega_\text{x}^2x^2 + \omega_\text{y}^2y^2 + \omega_\text{z}^2z^2)/2$ with $\omega_\text{x}=\omega_\text{y}=\gamma\omega_\text{z}$ and $\gamma\gg 1$. The tight radial trapping indicates that the excitation to the higher radial vibrational states can be neglected, and we assume in our calculations that all atoms populate the lowest radial eigenstate. The wavefunction can be approximately written as $\phi(r,t)=\left(\frac{m\omega_\text{x}}{\pi\hbar}\right)^{1/2}e^{-m\omega_\text{x}(x^2+y^2)/2\hbar}\phi(z,t)$, substituting which into the equations above and integrating along the radial directions we get the coupled equations determining the dynamics of the wavefunction $\phi$ and Bogoliubov amplitude along the $z$ direction. The reduced equations have the same forms with the Eqs.(\ref{eq2})-(\ref{eq4}) but with $H_0$ and $g$ substituted by $H_{0z} = -\frac{\hbar^2}{2m}\frac{d^2}{dz^2}+\frac{1}{2}m\omega_\text{z}z^2$ and $g_{1d}=g{\gamma}/{2\pi}$ respectively. 

\section{comparison and parametric analysis}

\begin{figure}[]
\includegraphics[width= 0.5\textwidth]{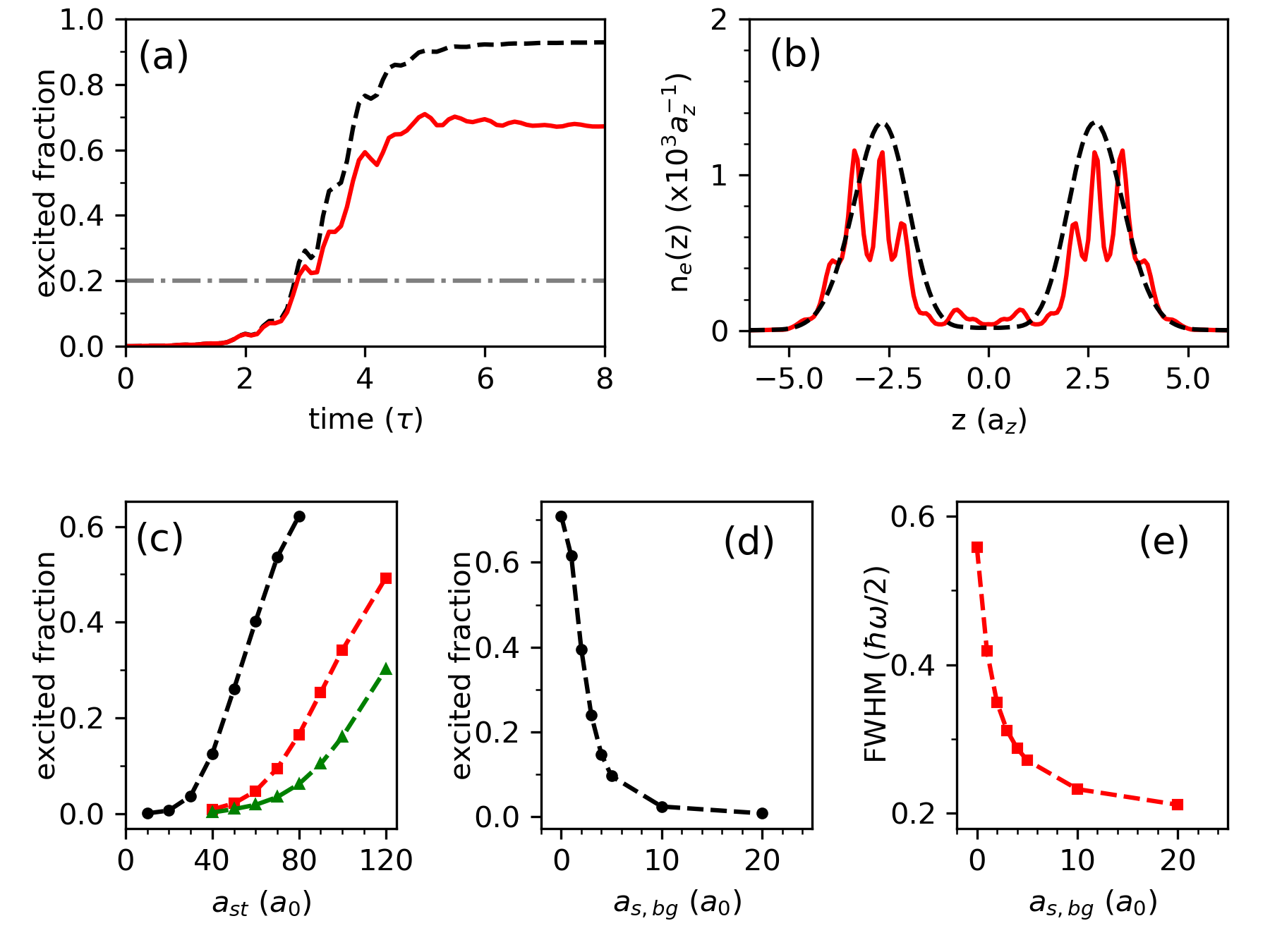}
\caption{\label{fig2} Stimulated excitation in a quasi one dimensional BEC. The trap frequency $\omega_\text{x}=\omega_\text{y}=2\pi\times 500~\text{Hz}$, $\omega_\text{z}=2\pi\times 20~\text{Hz}$. The total atom number $N=5000$, and the modulating frequency $\omega=2\pi\times 2000~\text{Hz}$. (a) shows the time dependence of the excited fraction, $N_\text{e}/N$, with typical Bogoliubov theory (black dash line) and td-GHFB theory (red solid line) respectively. The horizontal dotted dash line indicates a critical excited fraction where the contributions from the excited atoms become significant. The time is in unit of the modulation period $\tau=2\pi/\omega$. (b) shows the spatial density distributions of the excited atoms calculated from the typical Bogoliubov theory (black dash line) and the td-GHFB theory (red solid line) respectively with a modulating time of $8\tau$. Here $a_z = \sqrt{\hbar/m\omega_z}$. (c) and (d) show the dependences of the excited fraction on the modulating amplitude and the background scattering length respectively. In (c), the background $a_\text{s,bg}=0$ (black circle), 5$a_0$ (red square), 10$a_0$ (green triangle) respectively, while in (d), the amplitude $a_\text{st}=30a_0$. (e) gives the FWHM of the spectra distribution of the excited modes shown in Fig.\ref{fig1}(c) as a function of the background $a_\text{s,bg}$. In (c)-(e), a short modulating time of $2\tau$ is used and the dashed lines are just guides to eyes.}
\end{figure}

We first make a comparison between the results from the td-GHFB model and those obtained with the typically-used time-dependent Bogoliubov theory. As shown in Fig.\ref{fig2}(a), the growth trends of the excited fraction ($N_\text{e}/N$ with $N_\text{e}(t)=\int n_\text{e}(z,t)dz$) from the both models share an analogous shape, that is, initially growing exponentially while finally reaching a saturated value, which agrees with the observation in the 2D Bose-fireworks experiment \cite{Clark2017}. However, as shown clearly, the saturated excited fraction calculated with the td-GHFB theory is relatively much smaller, and the growth becomes slower after the excited modes are largely populated. This indicates the necessity of including the interacting contributions from the excitations, although, not surprisingly, during the early stage of the modulation the trends of the increase from the two models nearly coincide with each other until the excited population $N_\text{e}$ reaches a critical value of about $0.2N$. Another issue we are interested in is the spatial distribution of the excited population. Figure \ref{fig2}(b) shows the distribution of the excited density $n_\text{e}(z)$ after a sufficient modulating time. With the typical Bogoliubov theory, the excited atoms get spatially separated and show a smooth two-peak distribution. When adding the corrections, the distribution shows a multi-peak shape, but the center positions of the envelope still locate near those obtained from the typical Bogoliubov theory. 

Next let us check the dependence of the excitations on the scattering length. Figure \ref{fig2}(c) shows the total excited fractions for various modulating amplitudes $a_\text{st}$ after a fixed short-time modulation, for example, two modulating periods. For the case with $a_\text{s,bg}=0$, the excitation is rather weak for small $a_\text{st}< a_\text{crit} (\approx20 a_0)$, but gets burst when the modulation becomes stronger. For strong modulations with $a_\text{st}\ge 80 a_0$, such a short-time modulation can make the excited fraction approach the saturated value. Such a tendency agrees well with the function $N_\text{e}=Aa^2_\text{st}+B(a_\text{st}-a_\text{crit})\Theta(a_\text{st}-a_\text{crit})$ [$\Theta(x)$ is the step function] used to model the measured data in the Bose-fireworks experiment \cite{Clark2017}. It is worth to mention that the critical value of $a_\text{crit}$ not only depends on the modulation frequency $\omega$ \cite{Clark2017}, but also varies for different background scattering length. The results with finite $a_\text{s,bg}$ in Fig.\ref{fig2}(c) indicate that a larger $a_\text{s,bg}$ leads to a larger $a_\text{crit}$. 

We have also calculated the excited population with different initial background scattering length $a_\text{s,bg}$, as shown in Fig.\ref{fig2}(d). In previous experiments \cite{Clark2017,Feng2018}, the values of $a_\text{s,bg}$ were chosen to be less than $5 a_0$, whose contribution was dropped in the theoretical model in momentum space \cite{Clark2017}. However, here we claim that the stimulated excitation is sensitive to the background scattering length. From Fig.\ref{fig2}(d), with larger $a_\text{s,bg}$, significantly exciting the BEC becomes harder, i.e., more modulating periods are required. And interestingly, the distribution of the excited modes (see Fig.\ref{fig1}(c)) is sensitive to $a_\text{s,bg}$ as well. Larger background scattering length leads to a narrower spectra distribution, that is, a smaller FWHM, indicating that the excited atoms tend to become more focused in the resonant ($\omega_\text{e} = \omega/2$) mode. According to the viewpoints of Refs.\cite{Wu2018,Fu2018}, such a periodic modulation is a symmetry-breaking process. A stronger background interaction strength makes significant excitation of the BEC require much more external energy, i.e., larger modulating amplitude or longer enough modulating time, since the time interval during which the BEC suffers from a negative scattering length becomes shorter. Consequently, for a fixed time and a fixed modulating amplitude, larger $a_\text{s,bg}$ leads to less significant momentum non-conservation \cite{Wu2018}, thus a narrower FHWM. Another point should be claimed is that, because of the non-zero trap potential along $z$ direction, once the atoms fly to the trap edge, they should overcome the potential and thus the energy-resonant condition collapses. Consequently, the peak of the excited spectra would not perfectly locate at the resonant mode. A careful theoretical study on the case without the trap has been already performed in momentum space in Refs.\cite{Clark2017,Feng2018}. 

\section{dynamics of the stimulated excitation}

\begin{figure}[b]
\includegraphics[width= 0.5\textwidth]{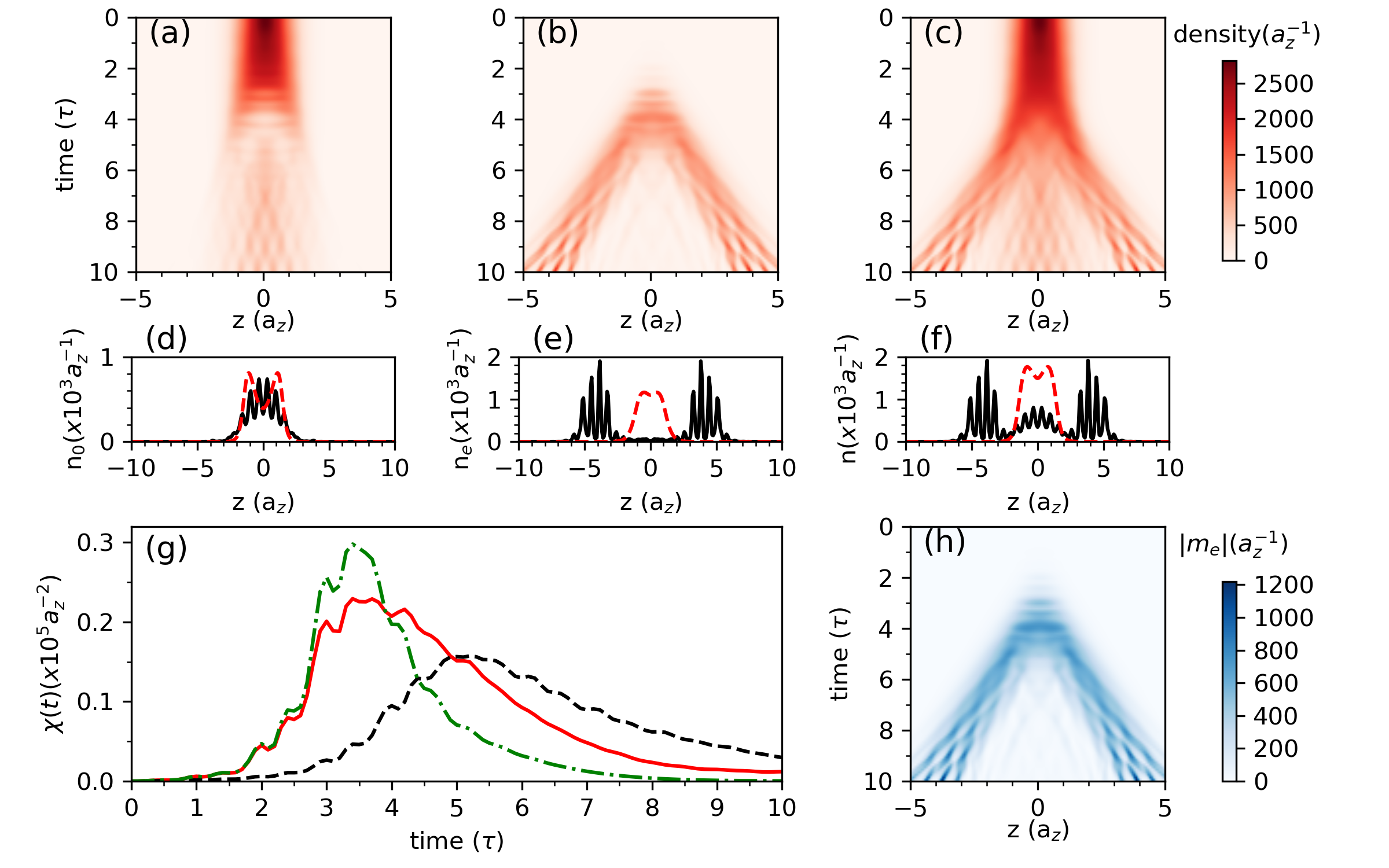}
\caption{\label{fig3} Dynamics of the stimulated excitation in a quasi one dimensional BEC. The trap frequency $\omega_\text{x}=\omega_\text{y}=2\pi\times 500~\text{Hz}$, $\omega_\text{z}=2\pi\times 20~\text{Hz}$. The total atom number $N=5000$, and the modulating frequency $\omega=2\pi\times 2000~\text{Hz}$ with the amplitude $a_\text{st}=30a_0$. For (a)-(f), $a_\text{s,bg}=0$. (a)-(c) show the time evolutions of the spatial density distributions of the condensate $n_0(z,t)$, the excitations $n_\text{e}(z,t)$ and the total $n(z,t)$, respectively. (d)-(f) respectively show the density profiles $n_0(z,t)$, $n_\text{e}(z,t)$ and $n(z,t)$ at finite modulating time $t=4\tau$(red dashed lines) and $t=10\tau$(black solid lines). (g) shows the time evolution of the defined density overlap $\chi(t)$. The green dotted dashed line is calculated from the typically-used Bogoliubov theory ($a_\text{s,bg}=0$), while the red solid line and black dashed line are calculated with td-GHFB theory for $a_\text{s,bg}=0$ and $a_\text{s,bg}=2a_0$ respectively. (h) gives the time evolution of the amplitude of the excited pairing field, $|m_\text{e}|$, which shows a similar behavior with the excited density profile in (b).}
\end{figure}

The td-GHFB theory provides an illuminating insight into the dynamics of the stimulated excitation. Figures \ref{fig3}(a) and (b) illustrate the time-evolutions of the spatial density distributions of the condensate $n_0(z)$, and the excitations $n_\text{e}(z)$, respectively. At the early stage of the modulation, $t\le 3\tau$, the excitation is weak and shows a distribution of a single-peak Gaussian shape, and the interactions among atoms in condensate dominate. The excitation totally overlaps with the condensate part as the size of the total density $n(z)=n_0(z)+n_\text{e}(z)$ distribution along $z$ does not get broadened yet, as shown in Fig.\ref{fig3}(c). Then, the excited population increases to be quantitatively approximate to the condensate, the interaction from the excited part plays a role in the evolutions of both the G-P and Bogoliubov equations. The condensate part gets spatially separated first, and then the excitations. For example, at time $t=4\tau$, the spatial profile of the condensate suffers from a multiple peak-valley phenomenon, while the separation of the excited part just begins at this moment; see Fig.\ref{fig3}(d) and (e). If we neglect the interaction terms from the excitations and use the typical Bogoliubov theory instead, the spatial density distribution of the condensate would not get any separation or broadened yet. Next, the evolution shows a rather interesting behavior. For the condensate part, multiple peaks and valleys appear and the atoms start to accumulate to the center position again. However, for the excitations, the spatial separation gets larger and larger and multiple peaks appear and propagate symmetrically. With a modulating time of $10\tau$, the condensate and the excited modes are almost totally separated, as shown in Fig.\ref{fig3}(c) and (f). 

To analysis the evolution behaviors above, we introduce a time dependent density overlap function
\begin{equation}\label{eq5}
\chi(t) = L^{-1}\int_{-L/2}^{L/2} dz~n_0(z,t) n_\text{e}(z,t)
\end{equation}
with the integration length $L=80a_z$ in our calculation. As a result, the averaged mean-field interaction energy between condensate and excitations can be expressed as $\zeta(t) = g(t)\chi(t)$. The dynamics of both $n_0$ and $n_\text{e}$ are related to this interaction energy term, $\zeta(t)$. Take $n_0=|\phi|^2$ as an example, the decomposition $\partial_t n_0 = \phi \partial_t\phi^* + \phi^*\partial_t\phi$ along with Eq.(\ref{eq2}) clearly leads to an interaction term $gn_0n_\text{e}$. Since $g(t)$ is a sinusoidal oscillation, we focus on the time-evolution of the function $\chi(t)$, as shown in Fig.\ref{fig3}(g). Once the modulation begins, the excited fraction increases and so does the overlap $\chi$, which reaches to the maximum value and then decreases to near zero as the modulation continues. The maximum value usually locates at the time when the spatial separation of the excitations begins. After the excited population reaching the saturated value (see Fig.\ref{fig2}(a)), the condensate and the excited modes start to separate with each other (Fig.\ref{fig3}(c)), and consequently the value of $\chi$ drops down and finally approaches to zero at time $t=10\tau$. Different parameters lead to different line shapes of $\chi(t)$. As shown in Fig.\ref{fig3}(g), the overlap shows a much more rapid decrease with the typical Bogoliubov theory, while the speeds of both the increase and the decrease become slower with a finite background scattering length.

The behavior of $\chi(t)$ indicates that the stimulated excitation is initially ($t<3\tau$) determined by the interactions among the condensate, while finally ($t>10\tau$) the evolutions of the condensate and the stimulated excitation are dominated respectively by the interactions among themselves. During the time interval $3\tau<t<10\tau$, the oscillation of the averaged interactions $\zeta(t)$ between the condensate and excitations dominates the dynamics, as shown in Fig.\ref{fig3}(c). Due to the interference between the condensate and the excited modes, the distribution of the total density $n(z)$ have multiple peaks which propagates simultaneously to left and right, just like a mechanical wave excited by a center driven source. Here the excitation source is the modulation of the interaction strength and the propagation is a kind of Faraday density wave \cite{Staliunas2002,Engels2007}. Meanwhile, the multiple peaks are somewhat like interference fringes (Faraday patterns \cite{Nicolin2007}), which should be attributed to the anomalous term $m_e(z,t)=\langle \hat{\psi}_e(z,t)\hat{\psi}_e(z,t)\rangle$, which describes the excited pairing field. 

The dynamics of the $|m_\text{e}|$ are shown in Fig.\ref{fig3}(h), illustrating a synchronous behavior with the excited density profile (see Fig.\ref{fig3}(b)). For short modulating time, although the spatial separation does not occur, the pairing process already happens. Then, the distribution of the pairing field gets spatially separated, in turn making an influence on the effective potential for the time evolution of the condensate wavefunction $\phi$, and consequently resulting in a broadened multi-peak density distribution of the condensate part. We have checked in our calculations that the multiple peaks would disappear and only the spatial separation could be observed without the $n_e$ and $m_e$ terms included. Interestingly, calculation without the $m_e$ terms shows that the broadening of the distribution of the condensate does not appear yet, although the multiple peaks still exist. Till now, we have made a vivid description of the stimulated excitation with a viewpoint of density wave propagation, and the fringe patterns come from the anomalous excited pairing field in the td-GHFB theory.

\section{High-order collisions}

\begin{figure}[]
\includegraphics[width= 0.5\textwidth]{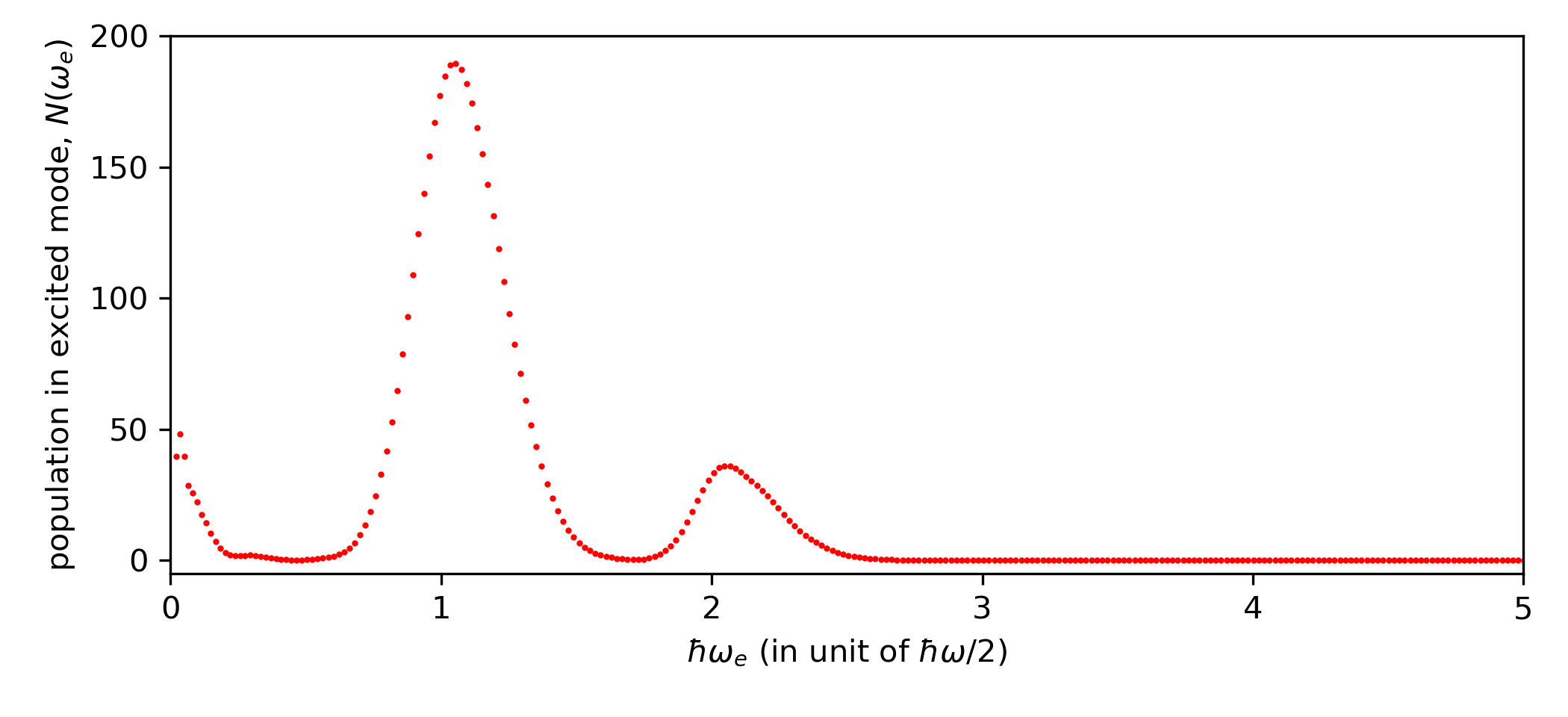}
\caption{\label{fig4} Generation of the second order matter-wave harmonic in a driven quasi one dimensional BEC. The trap frequency $\omega_\text{x}=\omega_\text{y}=2\pi\times 500~\text{Hz}$, $\omega_\text{z}=2\pi\times 20~\text{Hz}$. The total atom number $N=5000$, and the modulating frequency $\omega=2\pi\times 2000~\text{Hz}$. The background scattering length $a_\text{s,bg}=5a_0$ and the modulation amplitude $a_\text{st}=60a_0$. The modulating time is $10\tau$.}
\end{figure}

The td-GHFB theory can partly explain the high-order matter-wave harmonics observed in Ref.\cite{Feng2018}. The generation of the high-order harmonics depends on a strong enough modulation, and for the case in our calculation, the amplitude is usually larger than $50a_0$. Figure \ref{fig4} shows the excitation spectra, i.e., the distribution of the population in each mode, with the background scattering length $a_\text{s,bg}=5a_0$ and the modulation amplitude $a_\text{st}=60a_0$. Besides the modes with energy around $\hbar\omega_\text{e}=\hbar\omega/2$, another peak emerges with an energy of $\hbar\omega_\text{e}=\hbar\omega$, corresponding to the second order harmonic induced by the collisions between the multiple excited modes. However, other higher-order harmonics have not been observed yet even with a larger modulation amplitude or a longer modulation time. This is the restriction of the td-GHFB theory, although its performance is much better than the typical Bogoliubov theory where always only one peak at energy $\hbar\omega/2$ can be resolved. 

To show a clear picture of the restrictions, let us revisit the many-body Hamiltonian (\ref{eq1}). Here we do not  separate the condensate apart from the field operator $\hat{\psi}$, but expand it as $\hat{\psi} = \sum_k \phi_k \hat{a_k}$ in a spirit of Ref.\cite{Morgan2000}, where $\phi_0$ is calculated from the G-P equation, and $\phi_{k\neq 0}$ are determined by the Sch$\ddot{\text{o}}$dinger equation $[-\frac{\hbar^2\nabla^2}{2m}+\tilde{V}(r)]\phi_k = \epsilon_k\phi_k$ with $\tilde{V}(r) = V(r) + g|\phi_0(r)|^2$ to guarantee the orthogonality. Note that each eigen mode with energy $\epsilon_k$ ($k\ne 0$) is double degenerate in momentum space as the momentum can be either $k=\sqrt{2m\epsilon_k}/\hbar$ or $-k$. Now the Hamiltonian can be expressed as
\begin{equation}\label{eq6}
H = \sum_k \epsilon_k\hat{a}_k\hat{a}_k + \frac{1}{2}\sum_{p,q,k} G_{p+k,q-k,p,q} \hat{a}^\dag_{p+k}\hat{a}^\dag_{q-k}\hat{a}_p\hat{a}_q,
\end{equation}
where $G_{k,l,m,n} = g(t)\int d^3r \phi^*_{k}(r)\phi^*_{l}(r)\phi_{m}(r)\phi_{n}(r)$. Replacement of the creation operator $\hat{a}^\dag_0$ and annihilation operator $\hat{a}_0$ of the condensate by the $c$-number $\sqrt{N_0}$ leads to the typical Bogoliubov theory in momentum space, that is, the $\hat{a}^\dag_{k}\hat{a}^\dag_{-k}\hat{a}_0\hat{a}_0$ interaction term, which can directly explain the occurrence of the first peak in Fig.\ref{fig4}. The additional interaction contributions $n_e(z)$ and $m_e(z)$  from the excitations in Eqs.(\ref{eq2})-(\ref{eq4}) correspond to the $\hat{a}^\dag_{k'}\hat{a}^\dag_{-k'}\hat{a}_k\hat{a}_{-k}$ terms in the above Hamiltonian, which means that collision of an atom-pair with energy $\epsilon_k$ ($=\hbar\omega/2$, if considering the resonant condition) in the excitations happens by absorbing a external microwave photon with energy $\hbar\omega$, generating an atom-pair with energy $\epsilon_{k'}=\hbar\omega$ (i.e. $k'=\sqrt{2}k$). This is the physical picture of the collisions for the stimulated excitation. However, since the GHFB theory treats the non-quadratic terms with factorization approximations to reduce them to linear or quadratic terms  \cite{Morgan2000} and the condensate part is taken separately, higher-order collisional terms in Hamiltonian (\ref{eq6}), such as $\hat{a}^\dag_0\hat{a}^\dag_{k''}\hat{a}_k\hat{a}_k$, have been neglected, and consequently the td-GHFB theory could not explain the other higher harmonics observed in the experiment. But unfortunately, as far as we know, it is rather hard to directly solve the full Hamitionian (\ref{eq6}) numerically \cite{Morgan2000}, even for a one dimensional problem, which lies beyond the scope of our discussion here. The perturbation theory under some reasonable approximations used in Ref.\cite{Feng2018} might be a good choice to handle it.

\section{Conclusion}

In summary, the correction by adding the contributions of the interaction from the excited modes to the typical Bogoliubov theory leads to a strikingly different picture for the stimulated excitation in a driven BEC. A comparison and parametric analyses have been made. With the td-GHFB theory, the time-evolution of the density distribution shows a rather different behavior, that is, spatially separated with multi-peak-shape propagations. Meanwhile, the generation of the high order harmonics can be partly explained, although perfect explanation of the harmonics of all orders requires a full expansion of the time-dependent many-body Hamiltonian. Note that, in this work, for convenience, we have not considered the effect of thermalization \cite{Hu2018} during the increase of the excited fraction. Despite this problem, the td-GHFB theory can clearly account for both the density wave and the high-order collisions, illustrating a better performance than the typical Bogoliubov theory. The theoretical framework can be directly applied to future experiments with similar driven mechanism.

\begin{acknowledgments}
~

We acknowledge the support from the National Key Research and Development Program of China under Grant No.2018YFA0307200, National Natural Science Foundation of China under Grant No. 91636104, Natural Science Foundation of Zhejiang province under Grant No. LZ18A040001, and the Fundamental Research Funds for the Central Universities. 

\end{acknowledgments}

\bibliography{collective_excitation}

\end{document}